\documentclass[preprint, 5p, twocolumn]{elsarticle}

\usepackage{amsmath,amssymb,bm,multirow,float,bbold}
\usepackage[american]{babel}
\usepackage{flushend}
\allowdisplaybreaks

\newcommand{\lsim}{\stackrel{<}{_\sim}}

\newcommand{\be}{\begin{equation}}
\newcommand{\ee}{\end{equation}}

\usepackage{xcolor,colortbl}

\definecolor{Gray}{gray}{0.95}
\definecolor{RGray}{gray}{0.85}
\definecolor{CGray}{gray}{0.92}

\newcolumntype{a}{>{\columncolor{Gray}}c}
\newcolumntype{b}{>{\columncolor{white}}c}

\def\beqn{\begin{eqnarray}}
\def\ba{\begin{array}{c}}
\def\bat{\begin{array}{cc}}
\def\bat{\begin{array}{cc}}
\def\ea{\end{array}}
\def\bat{\begin{array}{cc}}
\def\batt{\begin{array}{ccc}}
\def\eeqn{\end{eqnarray}}

\begin{document}
\begin{frontmatter}

\title{An invisible axion model with controlled FCNCs at tree level}

\author[a]{Alejandro Celis}
\ead{alejandro.celis@ific.uv.es}

\author[a,b]{Javier Fuentes-Mart\'{\i}n}
\ead{javier.fuentes@ific.uv.es}

\author[a,c]{Hugo Ser\^odio}
\ead{hserodio@kaist.ac.kr}

\address[a]{Departament de F\'{\i}sica Te\`{o}rica and IFIC, Universitat de
Val\`{e}ncia-CSIC, E-46100, Burjassot, Spain}

\address[b]{Department of Physics, National Tsing Hua University, Hsinchu 300, Taiwan}

\address[c]{Department of Physics, Korea Advanced Institute of Science and Technology, 335 Gwahak-ro, Yuseong-gu, Daejeon 305-701, Korea}

\begin{abstract}
We derive the necessary conditions to build a class of invisible axion models with Flavor Changing Neutral Currents at tree-level controlled by the fermion mixing matrices and present an explicit model implementation.  A horizontal Peccei--Quinn symmetry provides a solution to the strong CP problem via the Peccei--Quinn mechanism and predicts a cold dark mater candidate, the invisible axion or familon.    The smallness of active neutrino masses can be explained via a type I seesaw mechanism, providing a dynamical origin for the heavy seesaw scale.   The possibility to avoid the domain wall problem stands as one of the most interesting features of the type of models considered.      Experimental limits relying on the axion-photon coupling, astrophysical considerations and familon searches in rare kaon and muon decays are discussed. 
\end{abstract}

\end{frontmatter}

\section{Introduction \label{s:intro}}

One of the most intriguing aspects of our current understanding of Nature is the remarkable success of the Cabibbo--Kobayashi--Maskawa (CKM) picture of the Standard Model (SM)~\cite{Cabibbo:1963yz} together with the stringent limits on the neutron electric dipole moment~\cite{Baker:2006ts}.       This is the root of the strong CP problem~\cite{Cheng:1987gp}.    The most compelling solution to this issue is probably the Peccei--Quinn (PQ) mechanism~\cite{Peccei:1977hh}, which predicts the existence of a very light and weakly coupled pseudo-Goldstone boson (the axion) resulting from the spontaneous breaking of an anomalous chiral $\mathrm{U(1)}_{\mbox{\scriptsize PQ}}$ symmetry~\cite{Weinberg:1977ma}.    

The historical development of axion models has been tightly related with the flavor problem of multi-Higgs-doublet models.    By enlarging the scalar sector of the SM with additional Higgs doublets, flavor changing neutral currents (FCNCs) appear in the scalar sector at tree-level. These currents are not suppressed in general, causing a disaster from the phenomenological point of view since scalar mediated flavor changing transitions are tightly constrained~\cite{Bona:2007vi}.  

The scalar sector of the original PQ axion model, where the PQ and the electroweak (EW) symmetries were spontaneously broken at the same scale, consisted of two Higgs doublets. It was natural to assume then that the same PQ symmetry responsible for solving the strong CP problem was also protecting the theory against dangerous flavor changing interactions.   This was achieved by an appropriate choice of the PQ charges, enforcing Natural Flavor Conservation (NFC)~\cite{Glashow:1976nt}.  Experimental searches for the axion soon ruled out this scenario.   The first attempts to save the PQ solution to the strong CP problem led to variant axion models in which the assumption of NFC was dropped with the objective of suppressing some specific axion couplings~\cite{Peccei:1986pn}, these kind of models were also ruled out by experimental data.     

It was realized then that the only way out for the PQ mechanism was to decouple the breaking of the PQ symmetry from the EW symmetry breaking.  This led to the current status of axion models in which the axion mass and couplings are suppressed by the PQ symmetry breaking scale, assumed to be much higher than the EW scale.   Axions arising from these models are termed invisible, stressing the fact that these avoid most of the experimental limits without problems.    The Kim--Shifman--Vainshtein--Zakharov (KSVZ)~\cite{Kim:1979if} and Dine--Fischler--Srednicki--Zhitnitsky (DFSZ)~\cite{Zhitnitsky:1980tq} models stand nowadays as benchmark invisible axion models. 

 We will consider in this letter invisible axion models of the DFSZ type where one only introduces additional scalar fields, keeping the same fermionic content as in the SM (with the possibility to include right-handed neutrinos). In the DFSZ model the PQ symmetry is used to enforce the NFC condition just as in the original PQ framework.    However, a PQ symmetry with a richer flavor structure is perfectly possible. Some of the advantages of a PQ symmetry that is not family universal but rather a horizontal symmetry were discussed in Ref.~\cite{Wilczek:1982rv}.   Explicit invisible axion models with a horizontal PQ symmetry can be found in Refs.~\cite{Davidson:1984ik}.  
 
Within the framework of multi-Higgs-doublet models, it has been realized that the NFC paradigm is not the only way to suppress dangerous flavor changing couplings in the scalar sector to phenomenologically acceptable levels.   In particular it has been pointed out that a flavor symmetry enforcing specific Yukawa textures can provide a satisfactory protection against dangerous flavor changing scalar interactions, even when FCNCs are present at tree level~\cite{Branco:1996bq}.  The protection is guaranteed in this case since the only sources of flavor changing phenomena are the fermion mixing matrices.        We determine in this letter the necessary conditions to build axion models with this feature. 

In the next section we briefly introduce the so-called Branco-Grimus-Lavoura (BGL) model~\cite{Branco:1996bq}, allowing us to put our motivations on a more solid ground. In Section~\ref{sec:2} we will prove that it is not possible to implement axion models within the two-Higgs-doublet model (2HDM) presented in Section~\ref{sec:1} and that an extension of the scalar sector is needed. In Section~\ref{sec:3}, we provide a simple extension of the BGL model for which the proposed mechanism is possible.   The most important features of our model as well as a discussion of the experimental constraints on the axion are presented in Section~\ref{sec:4}. We conclude in Section~\ref{sec:con}.

\section{The Branco-Grimus-Lavoura model} \label{sec:1} 
In what follows and in order to introduce some of our notation, we shall consider a 2HDM with the doublets denoted by $\Phi_j$ ($j=1,2$). Both Higgs doublets acquire a vacuum expectation value (vev) $\langle \Phi_j^0 \rangle = e^{i\alpha_j}v_j/\sqrt{2}$ with $ (v_1^2 + v_2^2  )^{1/2} \equiv v = (\sqrt{2}  G_F)^{-1/2}$ fixed by the massive gauge boson masses.  The presence of an additional Higgs doublet extends the Yukawa Lagrangian in both sectors, which now takes the general form
\be
-\mathcal{L}_{\mbox{\scriptsize{Y}}}=\overline{Q_L^0}\left[\Gamma_1\Phi_1 +\Gamma_2\Phi_2\right]d^0_R+\overline{Q_L^0}[\Delta_1 \widetilde{\Phi}_1+\Delta_2 \widetilde{\Phi}_2]u^0_R+\text{h.c.} \,,
\ee  
where $\tilde \Phi_j = i \sigma_2 \Phi_j^*$ with $\sigma_2$ being the Pauli matrix.  Here $Q_L^0$ stands for the left-handed quark doublets while $u^0_R$ and $d^0_R$ denote the right-handed  up and down quarks, all of them in a generic flavor basis. There are two independent complex Yukawa matrices in each sector, i.e. $\Delta_j$ and $\Gamma_j$. This will generically lead to flavor changing scalar interactions at tree-level, which implies severe phenomenological constraints~\cite{Bona:2007vi}. The canonical solution to this problem is enforcing NFC in the model, which in practical terms stands for requesting the simultaneous diagonalizability of the Yukawa matrices in each sector. In the 2HDM there are two standard ways of implementing the NFC condition:
\begin{itemize}
\item Through a symmetry, discrete or continuous, whose role is to restrict the number of Yukawas in each sector to one~\cite{Glashow:1976nt}.
\item Through Yukawa alignment. With this requirement the Yukawa matrices in the same sector have the same flavor structure up to an overall factor~\cite{Pich:2009sp}. 
\end{itemize}
The second condition is not implemented through a symmetry~\cite{Ferreira:2010xe}. However, it can be seen as an effective theory of a larger model with the first condition imposed at the UV level~\cite{Bae:2010ai}. Or even as a first order expansion in a minimal flavor violating scenario~\cite{Botella:2009pq,Buras:2010mh}.

Apart from these two flavor conserving solutions, another possibility is found in the BGL model~\cite{Branco:1996bq}. In this framework one of the sectors will have tree-level FCNCs, but these will be kept under control. This is achieved through some particular Yukawa textures
\beqn\label{eq:BGLTextures}
\begin{split}
\Gamma_1^{\mbox{\scriptsize{BGL}}}=&
\begin{pmatrix}
\times&\times&\times\\
\times&\times&\times\\
0&0&0
\end{pmatrix}, \,
\Gamma_2^{\mbox{\scriptsize{BGL}}}=
\begin{pmatrix}
0&0&0\\
0&0&0\\
\times&\times&\times
\end{pmatrix}\,, \\[0.3cm]
\Delta_1^{\mbox{\scriptsize{BGL}}}=&
\begin{pmatrix}
\times&\times&0\\
\times&\times&0\\
0&0&0
\end{pmatrix},\,\,
\Delta_2^{\mbox{\scriptsize{BGL}}}=
\begin{pmatrix}
0&0&0\\
0&0&0\\
0&0&\times
\end{pmatrix} \,, 
\end{split}
\eeqn
which can be easily implemented through an abelian symmetry.\footnote{This implementation is unique up to trivial permutations~\cite{Ferreira:2010ir,Serodio:2013gka}. Other variations can, nevertheless, give distinct phenomenological predictions~\cite{Botella:2014ska,Bhattacharyya:2014nja}.}  In this case the flavor matrices combinations (orthogonal to the mass matrices combinations) encoding the FCNCs take the following form in the fermion mass basis~\cite{Branco:1996bq}:\footnote{The matrices $N_d$ and $N_u$ are obtained by writing the Yukawa Lagrangian in the Higgs basis and diagonalizing the fermion mass matrices~\cite{Branco:1996bq}.}
\begin{align}\label{eq:NBGL}
\nonumber\left(N_{d}\right)^{\mbox{\scriptsize{BGL}}}_{ij}\;=\;&\frac{v_2}{v_1}\left(D_d\right)_{ij}-\left(\frac{v_2}{v_1}+\frac{v_1}{v_2}\right)(V^\dagger)_{i3}(V)_{3j}(D_d)_{jj} \,, \\
\left(N_u\right)^{\mbox{\scriptsize{BGL}}}\;=\;& \frac{v_2}{v_1}\text{diag}(m_u,m_c,0) -\frac{v_1}{v_2}\text{diag}(0,0,m_t)  \,.
\end{align}
The matrix $V$ represents the CKM quark mixing matrix and $D_{u,d}$ are diagonal quark mass matrices. The down-quark sector is the only one with tree-level FCNCs in this implementation. However, these are highly suppressed by the down-type quark masses and by off-diagonal CKM matrix elements. Detailed phenomenological studies of the BGL framework have been performed in Refs.~\cite{Botella:2014ska,Bhattacharyya:2014nja}. The BGL model presents several unique features, as explained above. However, it still suffers from a few problems. 

The first problem appears in the scalar sector of the model. While one may implement the desired Yukawa textures through an abelian discrete group, the scalar sector will always possess an accidental global $\mathrm{U(1)}$ symmetry~\cite{Branco:1996bq} which, after spontaneous symmetry breaking, introduces a Goldstone boson in the model. The addition of soft breaking terms or the inclusion of additional scalar singlets have been presented as alternatives to evade this problem~\cite{Branco:1996bq}. 

The second problem is related with the strong CP phase.  While there are no large contributions to electric dipole moments in the BGL model~\cite{Botella:2012ab}, this is based on the assumption of a vanishing or very small $\overline{\theta}$ term~\cite{Cheng:1987gp}.  

In this work we suggest that a global flavored PQ symmetry could be responsible for implementing the BGL Yukawa textures as well as solving the strong CP problem. This type of strong CP solution introduces a new particle, the axion, and bring several new features to the model. For instance, the axion is a well motivated cold dark matter candidate given that it is a very light and weakly interacting particle~\cite{Preskill:1982cy}.

\section{The anomalous condition for a BGL-type model}  \label{sec:2}

For the desired symmetry to be of the PQ type it has to be chiral and $\mathrm{SU(3)}_C$ anomalous. In what follows we shall find the necessary charge conditions for this scenario. We are then interested in finding the abelian generators such that
\vspace{-0.1cm}
\be
Q_L^0\rightarrow \mathcal{S}_L\,Q_L^0\,,\quad d_R^0\rightarrow \mathcal{S}_R^d\, d_R^0\,,\quad u_R^0\rightarrow \mathcal{S}_R^u \,u_R^0\,,
\ee
with
\vspace{-0.2cm}
\begin{align}
\begin{split}
\mathcal{S}_L=&\text{diag}(e^{iX_{uL}\theta},e^{iX_{cL}\theta},e^{iX_{tL}\theta})\,, \\ 
\mathcal{S}_{R}^d=&\text{diag}(e^{iX_{dR}\theta},e^{iX_{sR}\theta},e^{iX_{bR}\theta})\,, \\
\mathcal{S}_R^u=&\text{diag}(e^{iX_{uR}\theta},e^{iX_{cR}\theta},e^{iX_{tR}\theta})\,,\\
\end{split}
\end{align}
and
\be\label{eq:PhiTrans}
\Phi\rightarrow \mathcal{S}_\Phi\, \Phi\,,  \qquad \text{with} \qquad \mathcal{S}_\Phi=\text{diag}(e^{iX_{\Phi 1}\theta},e^{iX_{\Phi 2}\theta}) \,.
\ee
The Yukawa textures are dictated by the way the fermion fields transform, the Higgs field transformations will simply select a specific texture among all the possible ones~\cite{Botella:2009pq,Ferreira:2010ir,Serodio:2013gka,Botella:2012ab}. From Eq.~\eqref{eq:BGLTextures} we see that the Yukawas $\Delta_j$ are block diagonal in the up-charm sector. This forces the left- and right-handed up symmetry generators to be doubly degenerate, i.e.
\vspace{-0.1cm}
\begin{align}\label{eq:chargesaBGL1}
\begin{split}
\mathcal{S}_L&=\text{diag}\left(1,1,e^{iX_{tL}\,\theta}\right)\,, \\
\mathcal{S}^u_R&=\text{diag}\left(e^{iX_{uR}\,\theta},e^{iX_{uR}\,\theta},e^{iX_{tR}\,\theta}\right)\,,
\end{split}
\end{align}
where we have set one of the charges to zero using a global phase transformation. The conditions
\begin{equation} \label{eq:CA1}
X_{tL}\neq0\, \qquad \text{and} \qquad X_{uR}\neq X_{tR} \,, 
\end{equation}
should be satisfied in order not to increase the generators degeneracy. The phases appearing in the up-quark Yukawa term are
\vspace{-0.05cm}
\begin{equation}
\Theta_u=\theta
\begin{pmatrix}
X_{uR}&X_{uR}&X_{tR}\\
X_{uR}&X_{uR}&X_{tR}\\
X_{uR}-X_{tL}&X_{uR}-X_{tL}&X_{tR}-X_{tL}
\end{pmatrix}\,,
\end{equation}
with $\Theta_u$ called the up-quark phase transformation matrix. From it we see that the additional condition
\be\label{eq:CA2}
X_{tL}\neq-(X_{uR}-X_{tR}) \,,
\ee 
is necessary in order to guarantee two distinct block structures. The generators in Eq.~\eqref{eq:chargesaBGL1}, together with the conditions~\eqref{eq:CA1} and \eqref{eq:CA2} form the complete and minimal set of required conditions for the up BGL textures. In order to pick up the desired textures we just have to attribute the correct charges to the Higgs fields. We then choose the scalar charges
\vspace{-0.2cm}
\begin{equation}\label{Sup}
\vspace{-0.2cm}
\mathcal{S}^{\mbox{\scriptsize{up}}}_{\Phi}=\text{diag}\left(e^{iX_{uR}\,\theta},e^{i(X_{tR}-X_{tL})\,\theta}\right)\,.
\end{equation}
This choice associates $\tilde{\Phi}_j$ with the $\Delta_j$ of Eq.~\eqref{eq:BGLTextures}. 

We now turn to the down-quark sector. The left-handed transformation is the same, since it is shared by the two sectors. Concerning the right-handed generator for the textures $\Gamma_j$, see Eq.~\eqref{eq:BGLTextures}, it is constrained to have the form~\cite{Serodio:2013gka}
\be\label{eq:chargesaBGL2}
\mathcal{S}_R^d=e^{iX_{dR}\theta} \, \mathbb{1}\,.
\ee
Therefore, the down-quark phase transformation matrix is
\vspace{-0.05cm}
\be
\Theta_d=\theta
\begin{pmatrix}
X_{dR}&X_{dR}&X_{dR}\\
X_{dR}&X_{dR}&X_{dR}\\
X_{dR}-X_{tL}&X_{dR}-X_{tL}&X_{dR}-X_{tL}
\end{pmatrix}\,.
\ee
We now see that the Eqs.~\eqref{eq:chargesaBGL1} and~\eqref{eq:chargesaBGL2}, together with the first part of condition~\eqref{eq:CA1}, form the minimal set of required conditions for the down  BGL textures. To pick up the desired textures we need the scalar transformation
\be\label{Sdown}
\mathcal{S}_{\Phi}^{\mbox{\scriptsize{down}}}=\text{diag}\left(e^{-iX_{dR}\,\theta},e^{i(X_{tL}-X_{dR})\,\theta}\right) \,.
\ee
Note that the two scalar transformations in Eqs.~\eqref{Sup} and~\eqref{Sdown} do not generally coincide. In the original BGL formulation it is crucial that the Higgs doublet that couples to $\Gamma^{\mbox{\scriptsize{BGL}}}_1$ in the down sector, is the same that couples to $\Delta^{\mbox{\scriptsize{BGL}}}_1$ in the up sector. The other implementation would introduce new textures, spoiling the BGL-type suppression. This requirement leads to the additional charge constraint
\vspace{-0.05cm}
\be
\vspace{-0.1cm}
X_{dR}=-X_{uR} \,.
\ee
Since we introduced a chiral symmetry we get the anomaly free condition for the PQ symmetry with the QCD currents
\vspace{-0.05cm}
\be
\left[\mathrm{SU(3)}_C\right]^2\times \mathrm{U(1)}_{\mbox{\scriptsize{PQ}}}:\quad 2X_{tL}-X_{tR}+X_{uR}=0\,.
\ee
If this anomaly free condition is satisfied it makes both $\mathcal{S}_{\Phi}^{\mbox{\scriptsize{up}}}$ and $\mathcal{S}_{\Phi}^{\mbox{\scriptsize{down}}}$ equal. This, in turn, connects the down $\Gamma_2^{\mbox{\scriptsize{BGL}}}$ texture with the $\Delta_2^{\mbox{\scriptsize{BGL}}}$ texture, which just tell us that the two Higgs BGL implementation is anomaly free. 

Therefore, if we request an anomalous implementation of the BGL textures we need to extend the model.

\section{3HDM with a flavored PQ symmetry} \label{sec:3}
In the previous section we showed that the Yukawa textures of the BGL 2HDM cannot be imposed via a PQ symmetry.  A simple solution is to join both $\mathcal{S}_{\Phi}^{\mbox{\scriptsize{up}}}$ and $\mathcal{S}_{\Phi}^{\mbox{\scriptsize{down}}}$ into a single generator. In this way we extend the model to a three Higgs scenario with the scalar doublets $\Phi_k$ ($k=1,2,3$), with vevs $\langle \Phi^0_k \rangle = e^{i \alpha_k} v_k/\sqrt{2} $ satisfying $v \equiv (v_1^2+v_2^2+v_3^2)^{1/2} = (\sqrt{2} G_F)^{-1/2}$. 
Following the charge constraints found in the previous section we set for the quark fields the flavored PQ charges
\be
X_{tL}=-2\,,\, X_{uR}=\frac{5}{2}\,,\, X_{tR}=-\frac{1}{2}\,,\, X_{d R}=-\frac{5}{2}\,.
\ee
For the leptonic sector many possible implementations of the PQ symmetry are available either with Dirac or Majorana neutrinos. We shall focus on the last scenario, introducing two right-handed neutrino fields, $N_{Ri}$ ($i=1,2$). These two fields transform under the PQ symmetry with the same phase, $X_{NR}$. The charge transformation for this sector take the form
\vspace{-0.1cm}
\be
\vspace{-0.1cm}
X_{\tau L}=1\,,\quad X_{lR}=-1/2\,,\quad X_{NR}=1/2\,,
\ee
where we have defined $X_{eR}=X_{\mu R}=X_{\tau R}\equiv X_{lR}$ and we have set $X_{eL}=X_{\mu L}=0$, without loss of generality, just as we did in the quark sector.

Current experimental bounds exclude axions coming from a PQ symmetry that is broken at the EW scale~\cite{Cheng:1987gp}.  Viable axion models can be obtained if one decouples the PQ symmetry breaking from the breaking of the EW symmetry.    This can be achieved in a similar way as in the DFSZ and KSVZ invisible axion models, adding a complex scalar singlet which acquires a very large vev $\langle S  \rangle = e^{i \alpha_{\mbox{\tiny PQ}}} v_{\mbox{\tiny PQ}}/\sqrt{2}$, with $v_{\mbox{\tiny PQ}}  \gg v$. The field $S$ will transform under the PQ symmetry as
\vspace{-0.1cm}
\be
\vspace{-0.1cm}
S\rightarrow e^{i\theta}S\,.
\ee
The Yukawa Lagrangian will then take the form 
\begin{align}
\nonumber-\mathcal{L}_{\mbox{\scriptsize{Y}}}=&\overline{Q_L^0}\,\left[\Gamma_1\, \Phi_1+\Gamma_3\, \Phi_3\right]d^0_R +\overline{Q_L^0}\,[\Delta_1\, \widetilde{\Phi}_1+\Delta_2\, \widetilde{\Phi}_2] u^0_R\\
\nonumber &+\overline{L_L^0}\,\left[\Pi_2\, \Phi_2+\Pi_3\, \Phi_3\right]l^0_R + \overline{ L_L^0} \,\Sigma_3\, \widetilde \Phi_3 N^0_R \\
&+ \overline{ (N_R^0)^{c}} A N_R^0 S^* + \mathrm{h.c.}\,,
\end{align}
with $L_L^0$ being the left-handed lepton doublets and $l^0_R$ denoting the right-handed charged leptons, all of them in a generic flavor basis.  Here $A$ is a general $2\times 2$ complex symmetric matrix.

Finally, the Higgs charges are given by
\vspace{-0.1cm}
\be
\vspace{-0.1cm}
X_{\Phi1}=5/2\,,\quad X_{\Phi2}=3/2\,,\quad X_{\Phi3}=1/2 \,,
\ee
and the only allowed phase sensitive terms in the scalar potential are 
\be\label{eq:senspot}
(\Phi_1^{\dag} \Phi_2 )  S\,,\,(\Phi_1^{\dag} \Phi_3 )  S^{2}\,, \,   (\Phi_2^{\dag}  \Phi_3 ) S\,,\, (\Phi_2^\dag \Phi_1)(\Phi_2^\dag \Phi_3)\,.
\ee
In this three-Higgs-doublet model implementation we get the following Yukawa textures:
\begin{align}\label{eq:aBGL}
\begin{split}
\text{Down:}&\quad\Gamma_1=\Gamma_1^{\mbox{\scriptsize{BGL}}}\,,\, \Gamma_3=\Gamma_2^{\mbox{\scriptsize{BGL}}}\;;\\ \text{Up:}&\quad\Delta_1=\Delta_1^{\mbox{\scriptsize{BGL}}}\,,\, \Delta_2=\Delta_2^{\mbox{\scriptsize{BGL}}}\;;\\
\text{Charged leptons:}&\quad\Pi_2=\Gamma_2^{\mbox{\scriptsize{BGL}}}\,,\, \Pi_3=\Gamma_1^{\mbox{\scriptsize{BGL}}}\;;\\
\text{Dirac neutrinos:}&\quad \Sigma_3=\left.\Delta_1^{\mbox{\scriptsize{BGL}}}\right|_{3\!\!\!/} \,.
\end{split}
\end{align}
The matrix $\left.\Delta_1^{\mbox{\scriptsize{BGL}}}\right|_{3\!\!\!/}$ corresponds to the original BGL texture but with the third column removed. The Yukawa textures implemented in this framework are stable under renormalization group evolution~\cite{Botella:2011ne}.

This model possesses FCNCs in the down-quark sector controlled by the matrices
\vspace{-0.2cm}
\be
\vspace{-0.04cm}
\begin{array}{ll}
\left(N_d^{\prime}\right)_{ij}&=(D_d)_{ij}-\dfrac{v^2}{v^2_3}(V^\dagger)_{i3}(V)_{3j}(D_d)_{jj}\,,\\
\left(N_d\right)_{ij}&=\dfrac{v_2}{v_1}\left(D_d\right)_{ij}-\dfrac{v_2}{v_1}(V^\dagger)_{i3}(V)_{3j}(D_d)_{jj}\,,
\end{array}
\ee
 in the basis where the quarks are mass eigenstates.   In the anomaly free three Higgs BGL implementation, tree-level Higgs mediated $|\Delta S|=2$ processes suppressed by only $(V_{cd}^\ast V_{cs})^2\sim\lambda^2$ ($\lambda \simeq 0.225$) appear, requiring some of the neutral scalar fields of the theory to be heavy~\cite{Botella:2009pq}. Nonetheless, in this framework, just like in the BGL 2HDM, we have suppressions of the order $(V_{td}^\ast V_{ts})^2\sim\lambda^{10}$.

The model also possesses tree-level FCNCs in the charged lepton sector, although not as suppressed as in the quark sector. These will be completely controlled by the Pontecorvo--Maki--Nakagawa--Sakata
(PMNS) matrix~\cite{Pontecorvo:1957qd}.      The flavor matrices encoding the FCNC interactions among charged leptons take the form 
\begin{align}
\nonumber\left(N_e^{\prime}\right)_{ij}&=-\dfrac{(v_1^2+v_2^2)}{v^2_3}(D_e)_{ij}+\dfrac{v^2}{v^2_3}(U^\dagger)_{i3}(U)_{3j}(D_e)_{jj}\,,\\
\left(N_e\right)_{ij}&=-\dfrac{v_1}{v_2}(U^\dagger)_{i3}(U)_{3j}(D_e)_{jj}\,,
\end{align}
 in the fermion mass basis.   Here $U$ represents the PMNS mixing matrix and $D_e$ the diagonal charged lepton mass matrix.

The smallness of active neutrino masses is understood in this framework via a type I seesaw mechanism~\cite{Minkowski:1977sc} once the scalar singlet, $S$, gets a vev.  This way the PQ symmetry breaking scale provides a dynamical origin for the heavy seesaw scale~\cite{Chikashige:1980ui}.   The effective neutrino mass matrix is given by
\vspace{-0.2cm}
\be
m_\nu\simeq -\frac{v_3^{2}e^{i\left(\alpha_{\mbox{\tiny PQ}}-2\alpha_3 \right)}}{2\sqrt{2}v_{\mbox{\tiny PQ}}}\Sigma_3 A^{-1} \Sigma_3^T \,.
\ee
One active neutrino remains massless because $m_\nu$ is singular, fixing the size of the other two neutrino masses.  For a normal hierarchy:  $m_2=\sqrt{\Delta m^2_{21}}\simeq 9$~meV and $m_3=\sqrt{|\Delta m^2_{31}|}\simeq  50$~meV~\cite{Forero:2014bxa}.    An inverted hierarchy on the other hand implies two quasi-degenerate neutrinos: $m_1 \sim m_2 \simeq 50$~meV.

\section{Discussion on axion and Higgs properties}  \label{sec:4}

The anomalous $\mathrm{U(1)}_{\mbox{\scriptsize{PQ}}}$ symmetry in the model presented in the previous section is spontaneously broken by the vev of the singlet field $S$ at a very high scale.   Non-perturbative QCD effects induce a potential for the axion field $(a)$ which solves the strong CP problem via the PQ mechanism and gives a small mass to the axion.
The axion mass is given by~\cite{Weinberg:1977ma}:
\be
m_a \simeq \frac{f_\pi m_\pi |C_{ag}|}{v_{\mbox{\tiny PQ}}} \frac{z^{1/2}}{\left(1+z\right)} \simeq 6~\text{meV} \times \left( \frac{  10^9~\text{GeV}  }{ v_{\mbox{\tiny PQ}}/|C_{ag}| }  \right) \,,
\ee
and hence it becomes suppressed by the high PQ symmetry breaking scale.  Here $m_\pi \simeq 135$~MeV and $f_\pi \simeq 92$~MeV are the pion mass and decay constant, respectively.   The parameter $z$ is given by $z = m_u/m_d \simeq 0.56$.   The quantity $C_{ag}$ is determined by the chiral color anomaly of the current associated with the $\mathrm{U(1)}_{\mbox{\scriptsize{PQ}}}$ transformation~\cite{Adler:1969gk}, in our model it is given by
\vspace{-0.1cm}
\be \label{chiralcolor}
C_{ag}\equiv\sum_{i=\text{colored}}X_{iR}-X_{iL}=1  \,.
\ee
One of the interesting features of having a flavored PQ symmetry is that it is possible to avoid the formation of domain walls during the evolution of the Universe~\cite{Wilczek:1982rv,Zeldovich:1974uw}.  In our scenario the domain wall number is $N_{\mbox{\scriptsize{DW}}}=|C_{ag}| =1$, thus avoiding the domain wall problem~\cite{Barr:1986hs}.

\begin{figure}[ht!]
\centering
\includegraphics[width=0.5\textwidth]{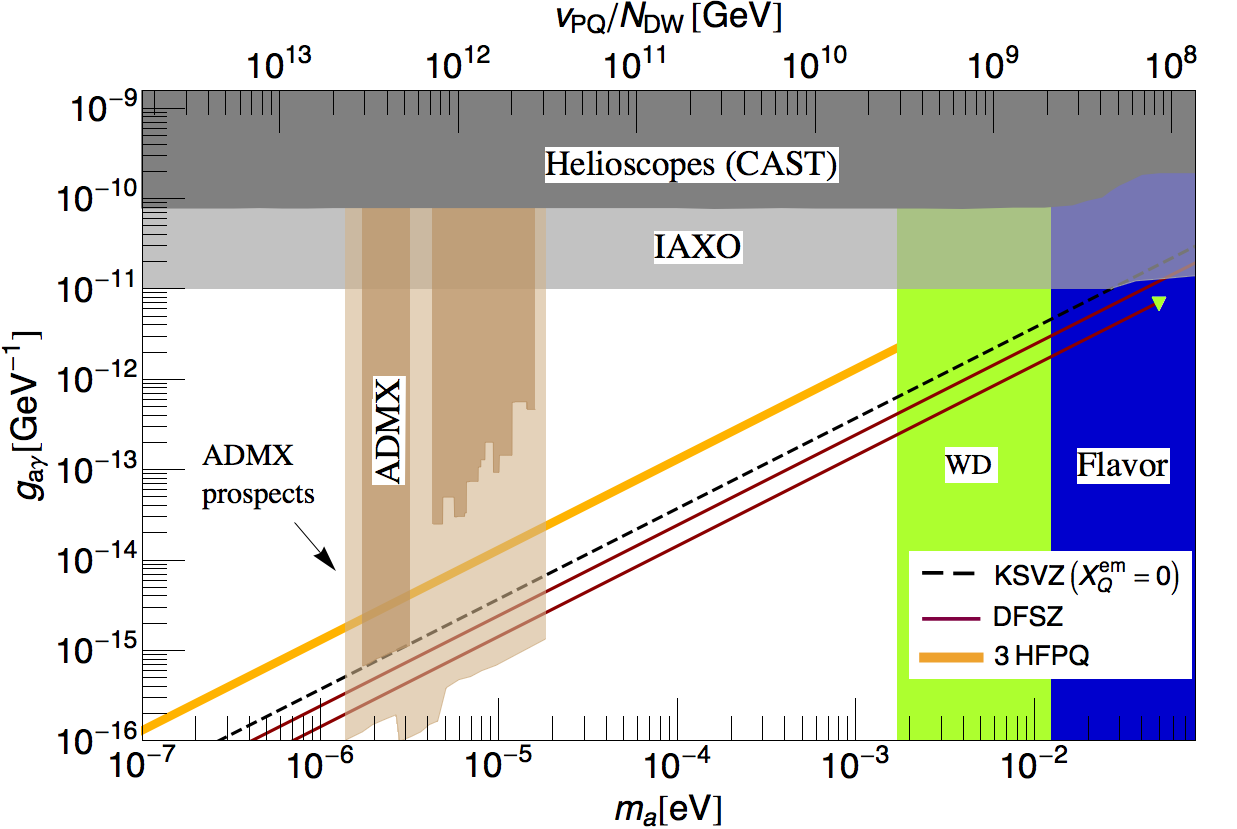}
\caption{\label{fig:constraints} \it \small     Constraints on the 3HFPQ axion from flavor experiments ($\mu^+ \rightarrow e^+ a \gamma $ and $K^+ \rightarrow \pi^+ a$), white-dwarfs (WD) cooling and axion-photon conversion experiments.      Predictions for the KSVZ (with $X_{Q}^{em}=0$) and DFSZ models are also shown.        The upper DFSZ band corresponds to the flipped scenario while the lower band to the type II.   The WD cooling constraint is shown at the benchmark point $v_2 \simeq v$ for the 3HFPQ model while for the type II DFSZ we take the most conservative bound (triangle).}   
\end{figure}

The main features of our three-Higgs flavored PQ (3HFPQ) framework are presented in Table~\ref{tab:features}, a comparison with the DFSZ and KSVZ invisible axion models is also done.   In the KSVZ model one adds to the SM particle content a complex scalar gauge singlet ($S$) together with a color triplet and $\mathrm{SU(2)}_L$ singlet heavy vector-like quark $(Q)$ with electric charge $X_{Q}^{em}$.     The SM fields carry no PQ charge in the KSVZ model.      In the DFSZ model one introduces an additional Higgs doublet and a complex scalar gauge singlet.    There are two possible implementations of NFC in the DFSZ model, the Higgs doublet coupling to $l_R$ can couple either to down-type quarks (type II) or to up-type quarks (flipped).     The most significant differences of our 3HFPQ framework with the usual benchmarks for invisible axion models are the presence of tree-level flavor changing axion couplings as well as large deviations from the axion coupling to photons.   
\begin{table}[h]
\begin{center}  
\doublerulesep 0.7pt \tabcolsep 0.06in
\begin{tabular}{r||c|c|c}   \rowcolor{RGray}
Models&KSVZ&DFSZ&3HFPQ\\     \rowcolor{Gray}
\hline 
BSM fields&$Q$+$S$&$\Phi_2$+$S$&$\Phi_{2}$+$\Phi_3$+$S$\\  
\multirow{2}{*}{PQ fields}&\multirow{2}{*}{$Q$, $S$ }&$q,\, l,\, \Phi_{1,2},\, S$ &$q,\, l,\, \Phi_{1,2,3},\, S$ \\    
&&(flavor blind)&(flavor sensitive)\\[0.1cm] \rowcolor{Gray}
$C_{a\gamma}/C_{ag}$&$6(X^{em}_Q)^2$&$2/3, 8/3$&$26/3$\\
CtM&No&Yes&Yes\\  \rowcolor{Gray}
FCAI& No&No& Yes\\
$N_{\mbox{\scriptsize{DW}}}$&1&$3, 6$&$1$\\
\hline
\end{tabular}
\caption{ \it \small  Main features of the 3HFPQ framework compared with the DFSZ and KSVZ invisible axion models.  CtM stands for Couplings to Matter and FCAI for Flavor Changing Axion Interactions, both at tree level.      The different values for $C_{a\gamma}/C_{ag}$ and $N_{\mbox{\scriptsize{DW}}}$ in the DFSZ model correspond to different implementations of the PQ symmetry.    \label{tab:features} }
\end{center}
\end{table}
The axion coupling to photons is described by the Lagrangian
\be
 \frac{\alpha}{8\pi v_{\mbox{\tiny PQ}}} C_{ag} C_{a\gamma}^{\mbox{\footnotesize eff}} \,a \, F_{\mu\nu}  \widetilde{F}^{\mu\nu}  \equiv   \frac{1}{4} g_{a \gamma}  \,a \, F_{\mu\nu}  \widetilde{F}^{\mu\nu}   \,.
 \ee
Here $\alpha= e^2/4 \pi \simeq 1/137$, $F_{\mu \nu}$ is the electromagnetic field strength tensor and $\tilde F_{\mu \nu}$ its dual.   The factor $C_{a\gamma}^{\mbox{\footnotesize eff}}$ takes the form~\cite{Srednicki:1985xd}:
\be  \label{caf}
C_{a\gamma}^{\mbox{\footnotesize eff}} \simeq \frac{C_{a\gamma}}{C_{ag}}-\frac{2}{3}\frac{4+z}{1+z},
\ee
where the second term in $C_{a\gamma}^{\mbox{\footnotesize eff}}$ is a model independent quantity which comes from the mixing of the axion with the $\pi^0$ while $C_{a\gamma}$ and $C_{ag}$ are model dependent quantities associated to the axial anomaly.   In our model 
\begin{align}
\begin{split}
C_{a\gamma}=&2\sum_{i=\text{charged}}(X_{iR}-X_{iL})Q_i^2 = \frac{26}{3} \,,
\end{split}
\end{align}
while $C_{ag}$ was already introduced in Eq.~\eqref{chiralcolor}.  The parameter $g_{a\gamma}$ is known as the axion-photon coupling constant.     Limits on $g_{a\gamma}$ as a function of the axion mass are shown in Fig.~\ref{fig:constraints}.  The CERN Axion Solar Telescope (CAST)~\cite{Zioutas:2004hi} excludes large values of the axion-photon coupling constant in the range of axion masses considered. Expected limits from the International Axion Observatory (IAXO), a proposed fourth generation axion helioscope~\cite{Irastorza:2011gs}, are also shown.  Microwave cavity haloscopes, including the Axion Dark Matter experiment (ADMX), exclude a window for the dark matter axion around a few $\mu$eV.  The latest type of experiments searches for cold dark matter axions in the local galactic dark matter halo~\cite{DePanfilis:1987dk,Carosi:2013rla}.   Limits from massive stars~\cite{Carosi:2013rla,Friedland:2012hj}, though not explicitly shown, put a limit similar to that from CAST.

A pseudo-Goldstone boson arising from the spontaneous breaking of a horizontal symmetry, known as a familon, could be detected in kaon or muon decays~\cite{Reiss:1982sq,Wilczek:1982rv,Feng:1997tn}.  In our framework the flavor changing couplings are controlled by the fermion mixing matrices and a robust upper bound on the axion mass can be extracted from the experimental limits on these processes.   The relevant flavor violating interactions are described by
\be
 \frac{ \partial_{\mu}a }{2 v_{ \mbox{\scriptsize{PQ}}  }}  \Bigl[     \bar \mu \gamma^{\mu}  \left(   g_{\mu e}^{V}  + \gamma_5 \,g_{\mu e}^{A}  \right)e + \bar s \gamma^{\mu}  \left(   g_{sd}^{V} + \gamma_5\, g_{sd}^{A}  \right) d    \Bigr] + \mathrm{h.c.} \,,
\ee
with 
\be
g_{sd}^{V,A} = -2V_{ts}^* V_{td} \,, \qquad   g_{\mu e}^{V,A} = U_{\tau2}^* U_{\tau1}   \,.
\ee
A limit $m_a \leq 12$~meV is obtained from $\mu^+ \rightarrow e^+ a \gamma$ decays~\cite{Bolton:1988af} while $m_a \leq 18$~meV is derived from limits on $K^+ \rightarrow \pi^+ a$~\cite{Adler:2002hy}.\footnote{We do not use the experimental limits on $\mu^+ \rightarrow e^+ a$~\cite{Jodidio:1986mz} since these assume that the lepton flavor violating axion couplings are of vectorial type.  In our framework, vector and axial lepton flavor violating couplings are equal.}    These limits are also shown in Fig.~\ref{fig:constraints}, excluding axions heavier than $12$~meV in our 3HFPQ framework.        Future improvements on the $K^+\rightarrow \pi^+ a$ limits may be achieved at the NA62 experiment at CERN~\cite{Fantechi:2014hqa}.  Improving the limits on $\mu \rightarrow e a \gamma$ decays on the other hand is very challenging with the current experimental facilities, see discussion in Ref.~\cite{Hirsch:2009ee}.

Stellar evolution and white-dwarf cooling considerations give the strongest constraints on the axion coupling to electrons~\cite{Raffelt:2006cw,Bertolami:2014wua}. The axion-electron axial coupling is given in our model by
\be
g^A_{ee}  \, \frac{\partial_\mu a }{2v_{PQ}}\,\overline{e}\,\gamma^\mu \gamma_5\, e \,,\quad\text{with}\quad g_{ee}^A = -2 + |U_{\tau1}|^2 + \frac{v_2^2 + 2 v_3^2}{v^2}\,.
\ee
From the limit $|g_{ee}^{A}|  m_e/v_{\mbox{\scriptsize{PQ}}} <  1.3 \times 10^{-13}$~\cite{Raffelt:2006cw,Bertolami:2014wua}, we extract the mass bound
\be
m_a\lsim 1.5/|g_{ee}^A| \, \text{meV} \,,
\ee
which depends on the vevs of the scalar doublets.  Imposing a perturbativity bound on the top quark Yukawa, $|(\Delta_{2})_{33}| \lsim \sqrt{4 \pi}$, and scanning over the vevs we get the following range of variation for the axion-electron coupling: $|g_{ee}^A|\in [0, 1.8]$.   In the top-vev dominance regime, i.e. when $v_2\sim v$, one obtains an upper bound on the axion mass $m_a \lsim 1.7$~meV which puts the 3HFPQ axion below the expected sensitivity of the IAXO. The axion-nucleon interactions are constrained by the requirement that the neutrino signal of the supernova SN 1987A is not excessively shortened by axion losses~\cite{Raffelt:2006cw}.  We find these constraints to be similar to those coming from the bounds on the axion-electron coupling.  However, the SN 1987A limit involves many uncertainties which are difficult to quantify~\cite{Raffelt:2006cw}.    Fig.~\ref{fig:constraints} summarizes all the constraints on the axion discussed.  The axion can also be tested in dedicated laboratory experiments looking for oscillating nucleon electric dipole moments~\cite{Graham:2011qk,Stadnik:2013raa,Roberts:2014cga}, axion induced atomic transitions~\cite{Sikivie:2014lha} and oscillating parity- and time reversal-violating effects in atoms and molecules~\cite{Stadnik:2013raa,Roberts:2014cga}.

The model considered can also give rise to a rich phenomenology in the Higgs sector.  The scalar doublets will receive large corrections to the mass matrix coming from the PQ breaking scale. We have the diagonal contributions, i.e. $|\Phi_i|^2|S|^2$, and the off-diagonal ones present in Eq.~\eqref{eq:senspot}. Up to $\mathcal{O}(v^2/v_{PQ}^2)$ the mass-squared matrix for the doublets will have a democratic texture with an overall PQ scale. Therefore, a possible decoupling limit has the singlet and one doublet at the PQ scale, while the other two doublets remain at the weak scale. At the effective level, the model would then be very similar to the BGL 2HDM implementation.       A rich scalar sector at the weak scale opens the possibility to observe rare flavor transitions mediated by scalar bosons~\cite{Botella:2014ska} as well as the direct discovery of additional scalars at the LHC~\cite{Bhattacharyya:2014nja}.   Furthermore, correlations between the flavor structure of the Higgs sector and the axion couplings arise in our model due to the underlying PQ symmetry.  These correlations are experimentally testable in principle.

\section{Conclusions}  \label{sec:con}

We have derived the necessary conditions to build an invisible axion model with tree-level FCNC completely controlled by the fermion mixing matrices.    It was shown that these kind of models cannot be built with two Higgs doublets and we have provided an explicit implementation with three Higgs doublets.   

The invisible axion in our framework possesses flavor changing couplings at tree-level which are constrained experimentally from kaon and muon decay experiments.   Limits from rare flavor transitions, astrophysical considerations and axion searches relying on the axion-photon coupling were analyzed.   Astrophysical axion bounds depend in general on the vevs of the Higgs doublets, in some regions of the parameter space the bound obtained from white-dwarfs can be as strong as $m_a \lsim 0.8$~meV.   Flavor processes put an upper bound on the axion mass $m_a \lsim 12$~meV which does not depend on any free parameter of the model.  Future results from microwave cavity experiments are expected to probe our model for axions around $m_a \sim 1-20$~$\mu$eV.   

The axion or familon of this framework provides a well motivated cold dark matter candidate.  Moreover, it is possible to explain the smallness of active neutrino masses via a type I seesaw mechanism, providing a dynamical origin for the heavy seesaw scale.   We have also shown that the model studied in this letter has $N_{\mbox{\scriptsize DW}}=1$ and therefore avoids the domain wall problem. On the other hand, in this letter we are dealing with an ad-hoc PQ symmetry. It could be argued that the imposition of a symmetry which is anomalous and therefore broken at the quantum level is unnatural. Moreover, we have not discussed how to stabilize the PQ solution to the strong CP problem against gravitational effects, see Ref.~\cite{Dias:2014osa} and references therein. A possible solution to both issues can be achieved in models where the PQ symmetry is no longer ad hoc but the result of an underlying discrete gauge symmetry~\cite{Babu:2002ic}. The implementation of this mechanism in the model we presented lies beyond the scope of the present work and is addressed in another publication~\cite{Celis:2014jua}.

\section*{Acknowledgements}
The work of A.C. and J.F. has been supported in part by the Spanish MINECO and the ERDF from the EU Commission [Grants FPA2011-23778 and CSD2007-00042 (Consolider Project CPAN)]. J.F. also acknowledges VLC-CAMPUS for an ``Atracci\'o del Talent'' scholarship and the National Science Council of Taiwan for the ``Summer Program in Taiwan 2014 for Spanish Graduate Students''. Part of H.S.'s work was funded by the ERDF, Spanish MINECO, under the grant FPA2011-23596. H.S.'s work was also supported by the National Research Foundation of Korea (NRF)
grant funded by the Korea Government (MEST) (No.~2012R1A2A2A01045722), and also supported by Basic Science Research Program through the National Research Foundation of Korea (NRF) funded by the Ministry of Education, Science and Technology (No.~2013R1A1A1062597). Finally, H.S. acknowledges the Portuguese FCT project PTDC/FIS-NUC/0548/2012.

\end{document}